\documentclass[a4paper,11pt]{article}
\pdfoutput=1 

\usepackage{jheppub} 

\usepackage[T1]{fontenc} 

\title{\boldmath Palatini Higgs inflation and the refined dS conjecture}

\author{Yang Liu}

\affiliation[a]{Building 16, Unit 3, 601, Wangguanzhuang community 9th, Jinan, Shandong Province, China}

\emailAdd{xijubear2020@Outlook.com}

\abstract{The refined de Sitter derivative conjecture provides constraints to potentials that are low energy effective theories of quantum gravity. It can give direct bounds on inflationary scenarios and determine whether the theory is in the Landscape or the Swampland [1]. Any infationary model can be checked by these conditions and non-minimally coupled scalar field theory is not an exception. We consider the “Palatini Higgs inflation” scenario taking the refined de Sitter derivative conjecture into account. According to the latest cosmological observations from Planck 2018, BICEP2+$Keck$ [2,3,4] and the bound of non-minimal coupling parameter $\xi$ [5,6], we suggest that if the refined dS conjecture does indeed hold, then the Palatini Higgs inflation model cannot be the low energy effective theory of a consistent quantum gravity theory since the two parameters $c_1$ and $c_2$ are much smaller than order 1, which is inconsistent with the refined dS conjecture. Therefore we suggest that it is in the “Swampland”.}

\begin{document} 
\maketitle
\flushbottom

\section{Introduction}
\label{sec:intro}

Motivated by a number of string theoretical constructions with controlled approximations, it has recently been proposed in ref.[7] that the potential for scalar fields in string theory satisfies the bound [7,8],
\begin{equation}\label{eq:1.1}
|\nabla V| \geq \frac{c_1}{M_p} \cdot V,  
\end{equation}
where $c_1$ is a positive constant of the order of 1, and $M_p$ is the Planck mass.\\
The conjecture, which corresponds to the initial "dS derivative conjecture" had severe tension with regard to numerous phenomenological models, both in particle physics and inflation [1]. In particular, the conjecture rules out the de Sitter (dS) vacua of a consistent theory since $|\nabla V|= 0$ but $V>0$ for a dS vacuum, even if the dS vacuum is the vacuum energy solution of Einstein’s equation with a positive cosmological constant. Therefore, the cosmological constant scenario is excluded if this conjecture is assumed to be true [1]. In order to explain the currently observed universe, the quintessence field with an exponentially decaying potential can be adopted, namely [1],
\begin{equation}\label{eq:1.2}
V_Q (Q) = \Lambda^4_Q e^{-c_Q Q},  
\end{equation}
which allows the dynamical vacuum energy with the present values of the quintessence field, the scale parameter as $Q \sim 0$ and $\Lambda_Q$ of order $O(1)$ meV [1]. There have been many follow-up papers considering various implications of the conjecture, such as discussed in ref.[9-13].\\
Following the earlier version of the conjecture proposed by Obied, Ooguri, Spondyneikeo and Vafa [7], i.e., “dS derivative conjecture”, the latest version of the Swampland conjecture was recently proposed by Ooguri, Palti, Shiu and Vafa [8], namely, “the refined de Sitter (dS) conjecture”. This conjecture states that any scalar potential $V(\phi)$ for scalar fields in a low energy effective theory of a consistent quantum gravity must satisfy at least one of the following conditions: 
\begin{equation}\label{eq:1.3}
|\nabla V| \geq \frac{c_1}{M_p} \cdot V,  
\end{equation}
or
\begin{equation}\label{eq:1.4}
min(\nabla_i \nabla_j V) \leq -\frac{c_2}{M^2_p} \cdot V,
\end{equation}
where $c_1$ and $c_2$ are both positive constants of the order of 1 and the left-hand side of eq.$(1.4)$ is the minimum eigenvalue of the Hessian $\nabla_i \nabla_j V$ in an orthonormal frame [8].\\
Let us review the two swampland conjectures briefly. We live in a universe where the vacuum energy is positive [7]. This can be realized by
having a scalar field potential $V$ with a local minimum at a positive value, leading to a meta-stable or stable de Sitter (dS) vacuum [7]. Despite many efforts in string theory community have been devoted to constructing de Sitter (dS) vacuum, there still exist a number of issues [7]. Therefore, it is fair to say that these scenarios have not yet been rigorously shown to be realized in string theory [7].\\
Considering the difficulties in obtaining dS-like vacua in string theory, it is natural to consider the alternative possibility where no dS vacua, not even meta-stable dS vacua, can exist in a consistent quantum theory of gravity [7]. One may conjecture $|\nabla V| > A$ for some constant $A$, which can lead to this exclusion [7]. However, supersymmetric vacua with flat directions provide counter-examples [7]. To avoid these counter-examples, one may consider allowing $A$ to depend on scalar fields $\phi$ in such a way that $A(\phi) \leq 0$ for supersymmetric vacua [7]. A simple choice would be $A(\phi) = c \cdot V(\phi)$ with positive constant $c$, since $V \leq 0$ in supersymmetric vacua [7]. Therefore, in ref.[7], the authors proposed that $|\nabla V| \geq c \cdot V$ ($M_p =1$), namely, eq.$(1.3)$ as a swampland criterion in any low energy theory of a consistent quantum theory of gravity [7]. This conjecture is the so-called “dS swampland conjecture”. The norm $|\nabla V| $ of the potential gradient is defined using the metric on field space in the kinetic term of the scalar fields [7].\\ 
Furthermore, in order to connect the “dS swampland conjecture” to another better established swampland condition, which is known as the distance conjecture [8], the authors of ref.[8] proposed the “refined dS swampland conjecture” which we have stated in the third paragraph. If $V$ is non-positive or $M_p \rightarrow \infty$, then the conjecture becomes trivial. This refined version still excludes (meta-)stable de Sitter vacua [8].\\
A multitude of coupling constants, such as the string coupling or the volume of extra dimensions can control effective descriptions of string theory [8]. All such couplings are dependent on field, and in particular, any weak couplings are associated with large distances in field space. The distance conjecture states that a scalar field can always be possible to change its value from its original point with an arbitrarily large amount of geodesic distance and that towers of light states with masses,
\begin{equation}\label{eq:1.5}
m \sim e^{-a\Delta \phi}
\end{equation}
descend from the ultraviolet if the change $\Delta \phi$ is trans-Planckian [8].\\   
The newly added bound is rather easily satisfied for a generic potential in low scale, $\triangle \phi \ll M_p$, since [1,14]
\begin{equation}\label{eq:1.6}
M^2_p \frac{\nabla_i \nabla_j V}{V} \sim -\frac{M^2_p}{(\Delta \phi)^2} \ll -c_2 \sim O(1),
\end{equation}
where $\Delta \phi$ is the change of field $\phi$. For convenience, we will call eq.$(1.3)$ and $(1.4)$ “Condition 1” and “Condition 2”, respectively. In this article, we will focus our interest on a non-minimally coupled scalar field theory known as the Palatini Higgs inflation scenario [5,6,15,16], which has not been examined so far. The goal is to find the constraints for constants $c_1$, $c_2$ which result from the Palatini Higgs inflation potential and to test the inflationary condition of this model.\\
The rest of this article is structured as follows: In section 2, we will briefly review the Palatini Higgs inflation scenario. In section 3, the range of swampland parameters $c_1$ and $c_2$ of Palatini Higgs inflation  will be  obtained. In section 4, we will reexamine the requirements of inflation for the refined dS conjecture. In section 5 we will present the conclusions.

\section{The Palatini Higgs inflation model}
In the usual formulation of Higgs inflation, the action is minimized with respect to the metric [16]. This procedure implicitly assumes the existence of a Levi-Civita connection depending on the metric tensor and the inclusion of a York-Hawking-Gibbons term ensuring the cancellation
of a total derivative term with no-vanishing variation at the boundary [16,17,18]. One could alternatively consider a Palatini formulation of gravity in which the metric tensor and the connection are treated as independent variables and no additional boundary term is required to obtain the equations of motion [16,19]. Roughly speaking, this formulation corresponds to assuming an ultraviolet completion involving different gravitational degrees of freedom [16].\\
In this article, we neglect the running of the parameter $\xi$ and apply unitary gauge for the Higgs field, $H= (0, h)^T/\sqrt{2}$, the relevant part of the Lagrangian then reads [5]:
\begin{equation}\label{eq:2.1}
L = -\frac{M^2_p + \xi h^2}{2} R + \frac{1}{2} (\partial_{\mu} h)^2 - \frac{\lambda}{4} h^4,
\end{equation}
where $\xi > 0$ is the strength of the non-minimal coupling of the Lagrangian to gravity.\\
We can perform a Weyl transformation of the metric to make the kinetic term of $h$ canonical, i.e. [5],
\begin{equation}\label{eq:2.2}
\hat{g}_{\mu\nu} = \Omega^2 g_{\mu\nu},
\end{equation}
where
\begin{equation}\label{eq:2.3}
\Omega^2 = 1 + \frac{\xi h^2}{M^2_p},
\end{equation}
and the field can be redefined [5,20]
\begin{equation}\label{eq:2.4}
h = \frac{M_p}{\sqrt{\xi}} \sinh(\frac{\sqrt{\xi} \chi}{M_p}).
\end{equation}
Then the Lagrangian then becomes [5]
\begin{equation}\label{eq:2.5}
L = - \frac{M^2_p}{2} \hat{R} + \frac{1}{2} (\partial_{\mu} \chi)^2  - U(\chi),
\end{equation}
where the scalar potential energy is given by
\begin{equation}\label{eq:2.6}
U(\chi)= \frac{\lambda M^4_p}{4 \xi^2} \left( \tanh(\frac{\sqrt{\xi} \chi}{M_p}) \right)^4.
\end{equation}
Note that if we had used in the metric formulation of the theory, we would have obtained the same form of the Lagrangian but with a different potential $U (\chi)$ [5,20]. The potential of eq.$(2.6)$ gives rise to inflation at the field values $\sqrt{\xi} \chi \gg M_p$ [5,20]. More details can be found in ref.[16].\\
In the following sections, we will take $M_p = 1$. In other words, the inflationary regime is $\sqrt{\xi} \chi \gg 1$ and eqs.$(1.3)$ and $(1.4)$ become
\begin{equation}\label{eq:2.7}
|\nabla V| \geq c_1 \cdot V,
\end{equation}
\begin{equation}\label{eq:2.8}
min(\nabla_i \nabla_j V) \leq -c_2 \cdot V.
\end{equation}

\section{Swampland parameters of the Palatini Higgs inflation model}
Since there is only one scalar field $\chi$ in the Lagrangian $(2.5)$, it is convenient to define two functions:
\begin{equation}\label{eq:3.1}
F_1(\chi) =\frac{|dU(\chi)/d\chi|}{U(\chi)}, 
\end{equation}
and
\begin{equation}\label{eq:3.2}
F_2(\chi) =\frac{d^2 U(\chi)/d \chi^2}{U(\chi)}. 
\end{equation}
Then considering eq.$(2.6)$, we obtain
\begin{equation}\label{eq:3.3}
F_1(\chi) =\frac{4 [1-\tanh^2(\sqrt{\xi} \chi) ] \sqrt{\xi} }{\tanh(\sqrt{\xi} \chi)}, 
\end{equation}
and
\begin{equation}\label{eq:3.4}
F_2(\chi) =\frac{4 \xi [1-\tanh^2(\sqrt{\xi} \chi)] [3-5\tanh^2(\sqrt{\xi} \chi)] }{\tanh^2(\sqrt{\xi} \chi)}.  
\end{equation}
In the inflationary regime, $\sqrt{\xi} \chi \gg 1$, the value of $\tanh(\sqrt{\xi} \chi)$ is very close to 1 but less than 1, therefore $F_2(\chi)$ must be less than zero.\\
Next, we need to introduce two “slow-roll” field inflation indices [15,21,22]:
\begin{equation}\label{eq:3.5}
\epsilon_V = \frac{1}{2} \left(\frac{V'}{V} \right)^2, 
\end{equation}
\begin{equation}\label{eq:3.6}
\eta_V =  \frac{V''}{V}, 
\end{equation}
where $V$ is the potential energy in the Lagrangian. For the Palatini Higgs inflation model, we have
\begin{equation}\label{eq:3.7}
\epsilon_V = \frac{1}{2} F^2_1, 
\end{equation}
\begin{equation}\label{eq:3.8}
\eta_V = F_2. 
\end{equation}
In the slow-roll region, the tensor-to-scalar ratio $r$ and the spectral index $n_s$ can be expressed as [1]:
\begin{equation}\label{eq:3.9}
r= 16 \epsilon_V = 8 F^2_1,
\end{equation}
\begin{equation}\label{eq:3.10}
n_s = 1- 6 \epsilon_V + 2 \eta_V = 1-3F^2_1 + 2F_2,
\end{equation}
or 
\begin{equation}\label{eq:3.11}
F_1 = \sqrt{2 \epsilon_V} = \sqrt{\frac{r}{8}},
\end{equation}
\begin{equation}\label{eq:3.12}
F_2 = \eta_V = \frac{n_s - 1 + 3r/8}{2}.
\end{equation}
According to ref.[15], up to $O(1/\xi)$ corrections, the slow-roll parameters in the Palatini Higgs inflation take the form:
\begin{equation}\label{eq:3.13}
\epsilon_V \approx \frac{1}{8 \xi N^2_V},
\end{equation}
\begin{equation}\label{eq:3.14}
\eta_V \approx - \frac{1}{N_V},
\end{equation}
with
\begin{equation}\label{eq:3.15}
N_V = \int_{i}^{\chi} \frac{d\chi}{\sqrt{2 \epsilon_V}}
\end{equation}
defining the number of $e$-folds and $\chi$ the corresponding background field value.\\
In addition, up to order of $O(1/N^2_V)$, we have [15]:
\begin{equation}\label{eq:3.16}
n_s \approx 1- \frac{2}{N_V} + \frac{2.8}{N^2_V}.
\end{equation}
Here we will use the recent inflationary data of Cosmic Microwave Background radiation (CMB) from the Planck observatory [1,3,4] and also the data obtained by the BICEP2+$Keck$ CMB polarization experiments [1,2]:
\begin{equation}\label{eq:3.17}
n_s \simeq 0.965 \pm 0.004,  
\end{equation}
\begin{equation}\label{eq:3.18}
r \lesssim 0.06. 
\end{equation}
According to refs.[5,6], the bound of $\xi$ is
\begin{equation}\label{eq:3.19}
1.0 \times 10^6 < \xi < 6.8 \times 10^7. 
\end{equation}
In this article, we take $N_V = 50.9$ corresponding to $\xi = 10^7$ [5]. In fact, as can be seen from the following calculated results, the orders of magnitude of the results are more important than the specific values. Inserting $N_V = 50.9$ into eq.$(3.16)$, we obtain $n_s = 0.962$, which satisfies the lower bound of eq.$(3.17)$.\\
Combining eq.$(3.13)$, $(3.19)$ and using $N_V = 50.9$, we then obtain the bounds of $\epsilon_V$
\begin{equation}\label{eq:3.20}
7.095 \times 10^{-13} < \epsilon_V < 4.825 \times 10^{-11}, 
\end{equation}
From eq.$(3.9)$, $(3.11)$ and $(3.12)$, we can obtain the bounds of $r$ and $F_1$ as
\begin{equation}\label{eq:3.21}
1.135 \times 10^{-11} < r < 7.720 \times 10^{-10}, 
\end{equation} 
\begin{equation}\label{eq:3.22}
1.191 \times 10^{-6} < F_1 < 9.823 \times 10^{-6}, 
\end{equation}
and that
\begin{equation}\label{eq:3.23}
F_2 \approx -0.019.
\end{equation}
The bounds of $r$ eq.$(3.21)$ satisfy the observatory data eq.$(3.18)$, namely, $r \lesssim 0.06 $.\\
If the Palatini Higgs inflation model satisfies the refined dS conjecture, then
\begin{equation}\label{eq:3.24}
c_1 \leq 9.823 \times 10^{-6},
\end{equation}
or 
\begin{equation}\label{eq:3.25}
c_2 \leq 0.019.
\end{equation}
However, neither $c_1$ nor $c_2$ is positive constant of the order of 1 [1,8,14,21,22]. Therefore, we suggest that the Palatini Higgs inflation model cannot satisfy the refined dS conjecture. In other words, if the refined dS conjecture holds, then the Palatini Higgs inflation model cannot be the low energy effective theory of a consistent quantum gravity theory. In our analysis, the RG running of the non-minimal coupling $\xi$ has been neglected. In fact, this approximation is justified [5,6]. 

\section{Testing inflationary condition of the Palatini Higgs inflation model}
In this section, we will determine the value of $\sqrt{\xi} \chi$ and check if its value satisfies the inflationary condition, $\sqrt{\xi} \chi \gg 1$.\\
Firstly, we will use $F_1$ to check the condition. If $\xi = 6.8 \times 10^7$ and $F_1 = 1.191 \times 10^{-6}$ these values are inserted into eq.$(3.3)$, we then have
\begin{equation}\label{eq:4.1}
32984.845 m^2 + 1.191 \times 10^{-6} - 32984.845 = 0,
\end{equation}
where we have defined $m= \tanh(\sqrt{\xi} \chi) > 0$. We then derive that $m =\tanh(\sqrt{\xi} \chi) \approx 1$. Therefore, $\sqrt{\xi} \chi \gg 1$.\\
Similarly, if $\xi = 1.0 \times 10^6$ and $F_1 = 9.823 \times 10^{-6}$ and these values are inserted into eq.$(3.3)$, we then also have $\sqrt{\xi} \chi \gg 1$.\\
Secondly, we will use $F_2$ to check the condition. Based on eq.$(3.23)$, in the range $1.0 \times 10^6 < \xi < 6.8 \times 10^7$, the value of $F_2$ is always approximately -0.019.\\ 
If $\xi = 6.8 \times 10^7$ and these values are inserted into eq.$(3.4)$, we then have 
\begin{equation}\label{eq:4.2}
m^2_1 \approx 1,
\end{equation}
namely, $\sqrt{\xi} \chi \gg 1$. Or
\begin{equation}\label{eq:4.3}
m^2_2 \approx \frac{3}{5},
\end{equation} 
namely, $\sqrt{\xi} \chi  \approx 1.032$, which cannot satisfy the inflationary condition, $\sqrt{\xi} \chi \gg 1$. Therefore, for $\xi = 6.8 \times 10^7$, there is only one reasonable value of $\sqrt{\xi} \chi$.\\
Similarly, if $\xi = 1.0 \times 10^6$ and these values are inserted into eq.$(3.4)$, we have the only one reasonable solution $m \approx 1$, namely, $\sqrt{\xi} \chi \gg 1$.\\
Therefore, according to observational data, we can determine that the value of $\sqrt{\xi} \chi$ satisfies the inflationary condition, $\sqrt{\xi} \chi \gg 1$. In other words, the model can be consistent with observations. However, as we have pointed out in section 3, if the refined dS conjecture holds, then the Palatini Higgs inflation model cannot be the low energy effective theory of a consistent quantum gravity theory. 

\section{Conclusions and discussions}
The refined dS conjecture could provide important implications for the low energy effective theory models which may or may not be consistent with the quantum gravity theory [1].The conjecture suggests two related but independent conditions for any scalar potential $V (\phi)$ of a low energy effective theory for a consistent quantum gravity [7,8]:
\begin{equation}\label{eq:5.1}
|\nabla V| \geq c_1 \cdot V,
\end{equation}
or
\begin{equation}\label{eq:5.2}
min(\nabla_i \nabla_j V) \leq -c_2 \cdot V,
\end{equation}
which we have respectively called Condition 1 and Condition 2 in this article. The parameters $c_1$  and $c_2$ are both positive constants of the order of 1. In this paper, we closely examined a non-minimally coupled scalar field theory known as the Palatini Higgs inflation model as a concrete example  of a potentially realistic inflationary model and applied the refined dS conjecture to evaluate its consistency with  quantum gravity theory.\\
At first, we review the conclusions in section 3. On the one hand, we have defined two functions $F_1$ and $F_2$ to determine $\frac{|dU(\chi)/d\chi|}{U(\chi)}$ and $\frac{d^2 U(\chi)/d \chi^2}{U(\chi)}$ for Palatini Higgs inflation model, respectively. Then, we express $F_1$ and $F_2$ using the tensor-to-scalar ratio $r$ and the spectral index $n_s$ in the “slow-roll” regime. On the other hand, according to ref.[15], the spectral index $n_s$ in Palatini Higgs inflation model can be reexpressed by the number of $e$-folds $N_V$, i.e., eq.$(3.16)$. According to ref.[5], the bound of the coupling parameter $\xi$ in Palatini Higgs inflation model can be obtained. In this article, we take $N_V=50.9$ corresponding to $\xi = 10^7$ and obtain that $n_s = 0.962$ which is consistent with the CMB observational data [1,2]. In fact, as can be seen from the following calculated results, the orders of magnitude of the results are more important
than the specific values. Furthermore, we have applied the latest cosmological observations from Planck 2018, from BICEP2+$Keck$ [2,3,4] and the bound of $\xi$ [5,6] to determine the range of two functions $F_1$ and $F_2$. If the refined dS conjecture was indeed to hold, we derived the upper bounds of $c_1$ and $c_2$, namely, $c_1 \leq 9.823 \times 10^{-6}$ and $c_2 \leq 0.019$. However, the values of $c_1$ and $c_2$ are much smaller than $O(1)$. We theorefore suggest that if the refined dS conjecture was to hold, then the Palatini Higgs inflation model cannot be the low energy effective theory for a consistent quantum gravity theory.\\
We have also determined the value of $\sqrt{\xi} \chi$ using observational data in section 4. Using the upper and lower bound of $\xi$ and the corresponding $F_1$, we can find that $\sqrt{\xi} \chi \gg 1$. Since the value of $\sqrt{\xi} \chi$ satisfies the inflationary condition of Palatini Higgs inflation model, i.e., $\sqrt{\xi} \chi \gg 1$ [5,20], the model is self-consistent. We therefore suggest that the “Palatini Higgs inflation” scenario is in the “Swampland” if the refined dS swampland conjecture holds. \\
Our understanding of the “Swampland” is still very much in flux. Since it is not a mature field there are no underlying principles from which results can be rigorously derived. Nevertheless, the past few years have seen significant interest and work on the “Swampland” and as a result several ideas and proposals have emerged as strong candidates framed within a more coherent structure. This is encouraging and suggests that we  are on the right track towards uncovering some interesting and possibly impactful new physics. Of course, it is highly likely that the various conjectures will be modified, some perhaps will be ruled out altogether in view of counter-examples and others will be refined. There is also increasing evidence  that there are strong links between conjectures which were initially considered to be independent of each other, suggesting that we may eventually obtain a picture in which the conjectures are all consequences of a single underlying newly discovered principle [23].\\
In the future work, we will apply double field theory in order to investigate the “Swampland” scenario. Double field theory formalism allows a description of the fundamental string in a way that is manifestly symmetric under $T$-duality [24,25]. The cosmological consequences of the resulting low energy vacua have not yet been fully explored. The quest to understand how de Sitter space can emerge from both supergravity and string theory is also of crucial importance and  touches on aspects of the “Swampland” scenario.


\end{document}